\documentclass[12pt]{article}
\usepackage{graphicx}
\usepackage{latexsym}

\begin{document}
\setlength{\oddsidemargin}{0mm}
\setlength{\evensidemargin}{0mm}
\setlength{\textwidth}{16.cm}
\setlength{\topmargin}{10pt}
\setlength{\headheight}{10pt}
\setlength{\footskip}{36pt}
\setlength{\textheight}{20cm}
\pagenumbering{arabic}
%\pagenumbering{roman}
%\pagenumbering{Roman}
%\pagenumbering{alph}
%\pagenumbering{Alph}
%\setcounter{page}{1}

\setcounter{equation}{0}
\setcounter{figure}{0}
\def\mib#1{\mbox{\boldmath $#1$}}

\title{RR charges of D2-branes in group manifold and Hanany-Witten effect}

\author{Takahiro Kubota  $^{a)}$ \footnote{e-mail: kubota@het.phys.sci.osaka-u.ac.jp} and
Jian-Ge Zhou $^{b)}$ \footnote{e-mail: jgzhou@hep.itp.tuwien.ac.at} }

%\date{}

\maketitle

\begin{center}
a) {\it Department of Physics, Graduate School of Science, 
Osaka University, Toyonaka,  Osaka 560-0043, Japan }

b) {\it Institut f{\" u}r Theoretische Physik, 
Technische Universit{\" a}t Wien, 
Wiedner Hauptstrasse 8-10, A-1040 Wien, Austria  }

\end{center}

\begin{abstract}
By exploiting the correspondence between the Cardy boundary 
state in $SU(2)$ group manifold and the BPS D3-brane 
configuration in the full asymptotically flat geometry of 
NS5-branes, we show that the Hanany-Witten effect in 10D 
background is encoded in the Cardy boundary states. 
The two RR Page D0 charges of the $n$-th spherical D2-brane 
due to the contraction to $e$ or ($-e$) is interpreted, 
and attributed to the Hanany-Witten effect.
\end{abstract}

\vfill\eject

After the work of Bachas, Douglas and Schweigert \cite{bds} 
the issues of RR D0 charge and $U(1)$ flux quantization for 
the spherical D2-branes in $SU(2)$ group manifold have attracted 
much interest \cite{pawe}-\cite{figueroa}. In \cite{bds}, 
they calculated the RR charges of the spherical D2-branes 
via two approaches: one based on the Dirac-Born-Infeld action 
with WZ term; the other exploiting the exact 1-point functions 
in the boundary WZW model, and found that the RR charges of 
D2-branes are irrational in two calculations, but the $U(1)$ 
flux is quantized. In \cite{marolf1} and  \cite{marolf2}, it 
was pointed out that the Abelian Chern-Simons terms 
in the dynamics of the gauge fields, for instance, in massless 
type IIA supergravity, induce the modified Bianchi identity
\begin{eqnarray}
d\tilde{F}_{4}+F_{2}\wedge H_{3}=0,
\end{eqnarray}
where $\tilde{F}_{4}$, $F_{2}$, $H_{3}$ are gauge 
invariant field strengths 
 of rank 4, 2, 3, respectively, and make the definition of 
charge in a gauge theory be more subtle. At least there are 
three natural notions of charge in a theory with Chern-Simons 
term: `brane source charge', `Maxwell charge'  and `Page charge', 
and it is Page charge,  which corresponds to $U(1)$ flux, that  
should be quantized. 

In \cite{alekseev2} and  \cite{figueroa}, it was shown that 
in $SU(2)$ WZW model
there are two different RR Page D0 charges
\footnote{We adopt the notion as in \cite{marolf1} and 
\cite{marolf2}.
}  for the spherical D2-brane due to two contraction: 
either to $e$ or $(-e)$, and argued that the RR Page D0 charge 
should be only defined\footnote{
There is a difference for the definition of the RR Page D0 charge between 
\cite{alekseev2} and \cite{figueroa} by $C_{1}(TD)/2$.
} modulo $k$.  In \cite{alekseev2}, it was expected this new feature 
should contain some important dynamical information, probably 
related to the fact that the $H$-field belongs to a nontrivial 
cohomology class. 

In the present paper, we further explore the origin how 
two different RR Page D0 charges for D2-brane are induced 
from the 10D dynamical point of view.  In doing so, we consider 
the BPS D3-brane in the full asymptotically flat geometry of 
NS5-branes. Based on the numerical calculations in \cite{pelc}\footnote{Some
related calculations can also be found in \cite{lrs}.}, 
we observe there is one to one correspondence between the BPS 
D3-brane configuration with 
$z_{{\rm max}}\longrightarrow \infty $ and opening angle 
$\psi _{n}$, and the $n$-th Cardy boundary state in $SU(2)$ 
 group manifold. We construct the dual for the $n$-th Cardy 
boundary state, which corresponds to the BPS D3-brane configuration 
with $z_{{\rm max}}\longrightarrow -\infty $ and opening angle 
$(\pi - \psi _{n})$. As we show the dual Cardy boundary state 
$\vert k-n>_{C}$ can be obtained by rotating 
$\vert n >_{C}$ by $\pi $ in the plane $(y^{6}, y^{7})$, 
from the correspondence this rotating operation is topologically 
 equivalent to moving the upper D3-brane across to the other 
side of NS5-branes along $z$ direction. Since the BPS 
D3-brane configuration corrsponding to the $n$-th Cardy boundary 
state can be interpreted as that there are $n$ number of 
D1-strings which suspend between distant flat D3-brane and 
NS5-branes \cite{pelc}, by exploiting the correspondence between 
the Cardy boundary state and the D3-brane configuration in 
the full asymptotically flat geometry of the NS5-branes, we 
find that when the lower D3-brane passes through $k$ coincident 
NS5-branes, $k$ D-strings will be created, that is, the 
Hanany-Witten effect 
\cite{hananywitten}-\cite{callan1} in 10D curved background is 
encoded in the Cardy boundary states in $SU(2)$ group manifold.    
By rephrasing Hanany-Witten effect, we see that depending on which 
side the distant observer stays, he(she) will observe two sorts 
of the number of D-strings: either $n$ or $n-k$. As two 
different contraction to $e$ or $(-e)$ corresponds to the 
observer staying on the different side of NS5-branes from 
ten dimensional point of view, this explains why for the 
$n$-th spherical D2-brane (corresponding to the  $n$-th Cardy 
boundary state) one observes two kinds of RR Page D0 charges. 
However, in the above physical processes, the RR Page D0 charge 
is not conserved due to Hanany-Witten effect 
\cite{hananywitten}-\cite{callan1}.

Now let us recall the background fields around a stack of 
$k$ coinciding flat NS5-branes which is given by 
\cite{callan2} 
\begin{eqnarray}
ds^{2}&=&dx^{2} + f dy^{2},
\nonumber \\
e^{2\Phi }&=&g_{s}^{2}f,
\label{eq:2}
\\
H_{k\ell m}&=&-\epsilon _{k\ell mn}\partial _{n}f, 
\nonumber 
\end{eqnarray}
 where $\{ x^{\mu}\}=(x^{0}, x^{1}, \cdots \cdots x^{5})$ 
parameterize the directions along the NS5-branes, 
$\{y^{m}\}=(y^{6}, y^{7}, y^{8}, y^{9})$ are locations of 
the fivebranes. $\Phi $ and $H$ are the dilaton and NS 3-form 
field strength, and $g_{s}$ is the string coupling far from 
the branes. The harmonic function $f$ depends on the transverse 
space 
\begin{eqnarray}
f&=&1+\frac{k\ell _{s}^{2}}{r^{2}}
\nonumber \\
r&=&\vert {\mib y}\vert =\sqrt{k}\ell _{s}e^{\phi }. 
\label{eq:3}
\end{eqnarray}
The background (\ref{eq:2}) and (\ref{eq:3}) 
interpolates between Minkowski space with constant $\Phi $ 
and a vanishing $H$, and a near-horizon region in which the 
geometry is an asymptotically linear dilaton solution
\begin{eqnarray}
ds^{2}&=&dx^{2}+k\ell _{s}^{2}(d\phi ^{2}+d\Omega _{3}^{2}),
\nonumber \\
H&=&2k\ell _{s}^{2}\omega _{3}, 
\end{eqnarray}
where $d\Omega _{3}^{2}$ and $\omega _{3}$ are the metric 
and volume form on the unit 3-sphere $S^{3}_{6789}$, and it 
describes the geometry of a throat 
\begin{eqnarray}
R^{1, 5}\times R_{\phi }\times SU(2).
\end{eqnarray}
The CFT describing the $S^{3}_{6789}$ is the $SU(2)$ WZW model 
at level k. The $SU(2)$ group element $g$ is related to the 
coordinates on $S^{3}_{6789}$ via 
\begin{eqnarray}
g({\mib y})=\frac{1}{\vert {\mib y}\vert }
\{y^{6}{\bf 1}+i(y^{8}\sigma _{1}+y^{9}\sigma _{2}+y^{7}\sigma _{3})\}.
\end{eqnarray}
The $SO(4)\sim SU(2)_{L}\times SU(2)_{R}$ global symmetry 
corresponding to rotations in the 
$R^{4} (y^{6}, y^{7}, y^{8}, y^{9})$ acts on 
$g$ as $g\longrightarrow h_{L}gh_{R}$ where 
$h_{L(R)}\in SU(2)_{L(R)}$. 
Denoting the generators of $SU(2)_{L}$ ($SU(2)_{R}$) by 
$J^{a}$ ($\bar J ^{a}$), one finds that $J^{3}-\bar J^{3}$ 
generates rotations in ($y^{6}, y^{7}$) plane, while 
$J^{3}+\bar J^{3}$ is the generator of the rotations in 
($y^{8}, y^{9}$). For the following discussions, we choose 
the cylindrical coordinates $(z, \rho , \theta , \varphi )$
\begin{eqnarray}
(y^{6}, y^{7}, y^{8}, y^{9})
=(z, \rho {\rm cos}\theta , \rho {\rm sin}\theta {\rm cos} \varphi , 
\rho {\rm sin }\theta {\rm sin }\varphi )
\end{eqnarray}
and spherical coordinates to replace $(z, \rho )$ by
\begin{eqnarray}
(z, \rho )=(r{\rm cos}\psi , r{\rm sin}\psi )
\end{eqnarray}
where $\theta \in [0, \pi]$, $\varphi \sim \varphi +2\pi $, $\psi 
\in [0, \pi]$. 

To see what the physical significance of the Cardy boundary 
states \cite{cardy} is  in 10D supergravity background, 
we consider the BPS configuration in which the D3-brane is 
orthogonal to the NS5-branes, that is, preserving the 
$SO(3)_{789}$ symmetry. The typical feature for this BPS 
D3-brane configuration is that it includes an infinite tube 
which can be interpreted as D1-brane \cite{pelc}. The angle 
$\psi $ has a simple geometrical meaning: opening angle as 
shown in Fig.1

%%%%%%% Fig. 1 %%%%%%%%%%%%%%%%
\begin{figure}[htbp]
\begin{center}
\includegraphics*[scale=.40]{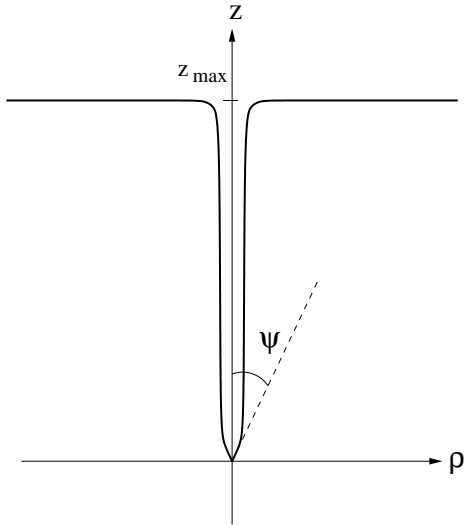}
\vspace{-0.7cm}
\end{center}
\caption{D3-brane profile for the fixed large 
$z_{{\rm max}}$ with opening angle $0< \psi < \pi /2$}
\end{figure}
%%%%%%%%%%%%%%%%%%%%%%%%%%%%%%%

\noindent
Especially when $z_{{\rm max}} \longrightarrow \infty $, the 
infinite tube can be identified with D1-branes \cite{pelc}. 

In $SU(2)$ group manifold, the Cardy boundary states 
are given by \cite{cardy}
\begin{eqnarray}
\vert n >_{C}=\sum _{j=0}^{k}\frac{S_{nj}}{\sqrt{S_{0j}}}
\vert j >_{I}
\end{eqnarray}
where $n=0, 1, \cdots , k$, $\vert j >_{I}$  is the Ishibashi 
state \cite{ishibashi} corrsponding to the chiral primary of 
spin $j/2$, and 
\begin{eqnarray}
S_{nj}=\sqrt{\frac{2}{k+2}}{\rm sin}\left (
\frac{(n+1)(j+1)\pi }{k+2}
\right )
\end{eqnarray}
is the modular-transformation matrix. The Cardy boundary state 
$\vert n >_{C}$ describes the $n$-th spherical D2-branes 
on $S^{3}_{6789}$ labelled by \cite{alekseev3} 
\begin{eqnarray}
\psi _{n}=\frac{n\pi }{k}
\end{eqnarray}
and possesses $n$ units of RR Page D0 charge which we shall 
interpret   below as the number of D1-branes in 10D curved 
background.

In the background of $k$ coincident NS5-branes, the $n$-th 
Cardy boundary state corresponds to D3-brane intersecting with 
 NS5-branes and taking the shape of a cone with opening angle 
$\psi _{n}$ in near-horizon region as illustrated in Fig. 2.

%%%%%%% Fig. 2 %%%%%%%%%%%%%%%%
\begin{figure}[htbp]
\begin{center}
\includegraphics*[scale=.55]{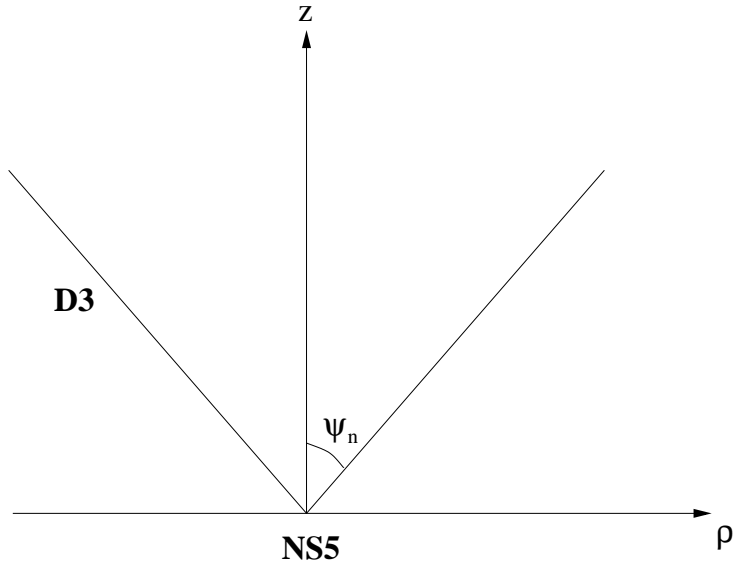}
\vspace{-0.7cm}
\end{center}
\caption{
The D3-brane intersects with the NS5-branes in the
 near-horizon region with the form of a cusp
}
\end{figure}
%%%%%%%%%%%%%%%%%%%%%%%%%%%%%%%

At the intersecting point ($z=\rho =0$), there is a singularity, 
thus the Dirac-Born-Infeld effective action is not an adequate 
 description of the test brane, however, in the present case 
of NS5-branes, we have an exact CFT in this region, 
with which we can do better, even work out Hanany-Witten effect 
which we shall show below. 

Fig. 2 only describes D3-brane in the near-horizon region, 
if we recall the full asymptotically flat geometry of the 
multiple NS5-branes, the D3-brane configuration should 
take the form in Fig. 1, which can be interpreted as 
D1-strings suspending between the flat D3-brane and the 
NS5-branes \cite{pelc}.

What we have seen from the above discussion is that based 
on the numerical calcultion in \cite{pelc} we observe that 
in 10D curved background, there is one to one correspondence 
between the $n$-th Cardy boundary state and the BPS D3-brane 
configuration with $z_{{\rm max}}\longrightarrow \infty $ 
and opening angle $\psi _{n}=n\pi /k$. The RR Page D0 charge 
of Cardy boundary state can be identified to the number of 
D1-branes which is depicted via the infinite throat \cite{pelc}.

We turn to construct the dual Cardy boundary state which 
corresponds to the D3-brane configuration with 
$z_{{\rm max}}\longrightarrow -\infty $ and opening angle 
$(\pi -\psi _{n})$ as drawn in Fig. 3. 

%%%%%%% Fig. 3 %%%%%%%%%%%%%%%%
\begin{figure}[htbp]
\begin{center}
\includegraphics*[scale=.4]{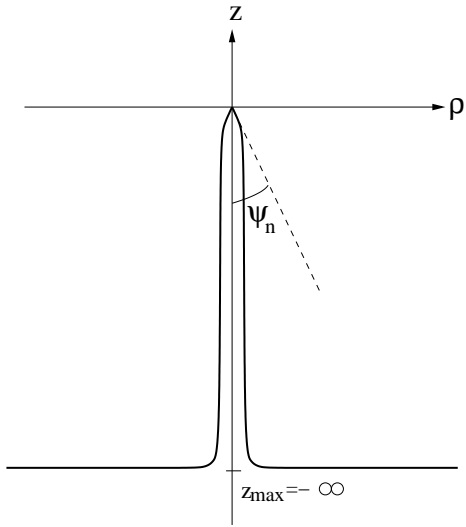}
\vspace{-0.7cm}
\end{center}
\caption{
The shape of D3-brane corresponding to the dual Cardy 
 boundary state
}
\end{figure}
%%%%%%%%%%%%%%%%%%%%%%%%%%%%%%%

\noindent
Since $J^{3}-\bar J^{3}$ and $J^{3}+\bar J^{3}$ are the 
generators of the rotations in $(y^{6}, y^{7})$ and 
$(y^{8}, y^{9})$ plane respectively, we construct the dual 
Cardy boundary state of $\vert n >_{C}$ by the rotation of 
$\pi $, i.e., 
$\vert n >_{C}^{D}={\rm exp} \{i\pi (J^{3}_{0}-\bar J^{3}_{0})\} 
\vert n >_{C}$. 
At first, we consider the bosonic case, and act the operator 
${\rm exp} \{i\pi (J^{3}_{0}-\bar J^{3}_{0})\}$ on Ishibashi 
state $\vert j>_{I}$ satisfying the boundary condition 
\cite{ishibashi} 
\begin{eqnarray}
(J_{n}^{3}+\bar J^{3}_{-n})\vert j >_{I}=0, 
\qquad 
(J_{n}^{\pm}+\bar J^{\pm}_{-n})\vert j >_{I}=0 
\label{eq:12}
\end{eqnarray}
which preserve maximal chiral Kac-Moody symmetry. One can show 
that the rotated boundary state 
${\rm exp}\{ i\pi (J^{3}_{0}-\bar J^{3}_{0}) \} \vert j > _{I}$ 
satisfies the same boundary condition (\ref{eq:12})
\footnote{
One can exploit the relation $e^{i\alpha J^{3}_{0}}J_{n}^{\pm }
e^{-i\alpha J^{3}_{0}}=e^{\pm i\alpha }J^{\pm }_{n}$.
}.
Since the general form of the Ishibashi state is 
\begin{eqnarray}
\vert j >_{I}=\Pi _{i} J^{a_{ i}}_{-n _{i}} \vert 0, j, j >
\otimes (-\bar J^{a _{i}}_{-n _{i}}) \vert 0, j, -j >
\end{eqnarray}
one can get 
\begin{eqnarray}
{\rm exp}\{ i\pi (J^{3}_{0}-\bar J^{3}_{0}) \} \vert j >_{I}
= (-1)^{j} \vert j >_{I}
\end{eqnarray}
and one has 
\begin{eqnarray}
\vert n >_{C}^{D}&=&\left (
\frac{2}{k+2}
\right )^{1/4}\sum _{j=0}^{k}\frac{{\rm sin}\left [
(n+1)(j+1)\pi/(k+2)\right ]}{
\sqrt{{\rm sin}\left [
(j+1)\pi /(k+2)
\right ]}}
(-1)^{j}\vert j >_{I}
\nonumber \\
&=&\vert k-n >_{C}
\end{eqnarray}
which indicates that the state corresponding to the D3-brane 
 configuration in Fig. 3 can be obtained by rotation operator 
${\rm exp}\{ i\pi (J^{3}_{0}-\bar J^{3}_{0})\} $ on the 
$n$-th boundary state which is consistent with the 
intuitive expectation. 

Since the boundary condition (\ref{eq:12}) preserves the 
diagonal $SU(2)$ symmetry $g\longrightarrow hgh^{-1}$, the 
Cardy boundary state $\vert k-n>_{C}$ can also be constructed 
by acting the other rotation operator 
${\rm exp}(2\pi i J^{3}_{0})$ (or 
${\rm exp}(-2\pi i \bar J^{3}_{0})$) on $\vert n >_{C}$, but 
with different bahavior in the supersymmetric generalization. 
The total $SU(2)$ current algebra of level $k+2$ is generated 
by the currents $J^{a}_{{\rm total}}=J^{a}+j^{a}$, 
where $j^{a}=-\frac{i}{2}\epsilon ^{abc}\psi _{b}\psi _{c}$ 
is the contribution of the fermions. After a certain chiral 
rotation, we have  
\begin{eqnarray}
j^{3}_{0}=\sum _{n \geq 1}^{\infty} \left (
\psi ^{1}_{n+1/2}\psi ^{2}_{-(n+1/2)}+
\psi ^{1}_{-(n+1/2)}\psi ^{2}_{n+1/2}
\right ), 
\nonumber \\
\bar j^{3}_{0}=-\sum _{n \geq 1}^{\infty} \left (
\tilde \psi ^{1}_{n+1/2} \tilde \psi ^{2}_{-(n+1/2)}+
\tilde \psi ^{1}_{-(n+1/2)} \tilde \psi ^{2}_{n+1/2}
\right )
\end{eqnarray}
for NS sector and 
\begin{eqnarray}
j^{3}_{0}=\psi ^{1}_{0}\psi ^{2}_{0}+\sum _{n \geq 1}^{\infty}
\left (\psi ^{1}_{n}\psi ^{2}_{-n}+\psi ^{1}_{-n}\psi ^{2}_{n}
\right ),
\nonumber \\
\bar j^{3}_{0}=- \tilde \psi ^{1}_{0} \tilde \psi ^{2}_{0}- 
\sum _{n \geq 1}^{\infty} \left (
\tilde \psi ^{1}_{n} \tilde \psi ^{2}_{-n}+\tilde \psi ^{1}_{-n} 
\tilde \psi ^{2}_{n}
\right )
\end{eqnarray}
for R sector, and the minus sigh between 
$j^{3}_{0}$ and $\bar j^{3}_{0}$ is due to the opposite chiral 
rotation. The boundary condition for the fermionic part is 
\begin{eqnarray}
(\psi ^{a}_{n} - i \eta \tilde \psi ^{a}_{-n})
\vert B_{\psi }, \eta >=0 
\end{eqnarray}
where $\eta =\pm 1$, $a=1,2,3$ and the fermionic boundary state 
takes the standard form as in flat space. Then the total boundary 
states of the $SU(2)$ WZW model factorize into
\begin{eqnarray}
\vert n, \pm > ^{{\rm susy}}=\vert n >_{C} \otimes \vert f, \pm >
\end{eqnarray}
with 
\begin{eqnarray}
\vert f, \pm > = \vert {\rm NS}\;  {\rm NS} > \pm \vert {\rm R}\;  
{\rm R}>
\end{eqnarray}
where $\vert f, \pm >$ denote the  fermionic boundary states 
for the brane and anti-brane. 

In the supersymmetric case, there are two sorts of the 
rotation operators given by 
\begin{eqnarray}
{\cal O}_{1}={\rm exp}\left [ i\pi (J^{3}_{0}-\bar J^{3}_{0})
+i\pi (j^{3}_{0}-\bar j^{3}_{0})\right ], 
\quad
{\cal O}_{2}={\rm exp}\left [ 
2\pi i (J^{3}_{0}+j^{3}_{0})\right ].
\end{eqnarray}
After the straightforward calculation, we have 
\begin{eqnarray}
{\cal O}_{1}\vert n, \pm >=\vert k-n, \pm >, 
\qquad 
{\cal O}_{2}\vert n, \pm >=\vert k-n, \mp >, 
\end{eqnarray}
which indicates that ${\cal O}_{1}$ operator retains (anti)brane 
as (anti)brane, while ${\cal O}_{2}$ operator reverses (anti) brane 
into (brane) anti-brane due to the fermionic zero mode in the RR 
sector which changes the sign of the boundary state 
$\vert {\rm R}\; {\rm R} >$. 

In \cite{alekseev2}, it was argued that there should exist a 
finite translation operator which moves D0-branes from $e$ to $(-e)$ 
and maps the Cardy boundary state $\vert 0, +>^{{\rm susy}}$ to 
$\vert k/2, ->^{{\rm susy}}$. Here we have explicitly constructed 
the operator ${\cal O}_{2}$ which maps $\vert n, \pm >^{{\rm susy}}$ 
to $\vert k-n, \mp >^{{\rm susy}}$, thus our construction 
confirms their conjecture.

We consider the pair of the Cardy boundary states $\vert n>_{C}$ 
and  $ \vert k-n >_{C}$, whose  RR Page D0 charges are $n$ 
and $k-n$. In Fig.4, the D3-brane configuration  
corresponding to the pair of Cardy boundary states   
$\vert n>_{C}$ and  $ \vert k-n >_{C}$ is portrayed\footnote{In Fig.4, we
do not mean we treat with the bound state by gluing together two BPS D3-brane
solutions (corresponding to the Cardy boundary states $\vert n>_{C}$ 
and  $ \vert k-n >_{C}$). Here we mean that when $t\rightarrow -\infty$, the
D3-brane locates at $z_{max}\rightarrow -\infty$, and when  $t\rightarrow
\infty$, the lower D3-brane after moving across the NS5-branes finally stays
at $z_{max}\rightarrow\infty$.}.

%%%%%%%%%%%%% Fig. 4 %%%%%%%%%%%%%%
\begin{figure}[htbp]
\begin{center}
\includegraphics*[scale=.40]{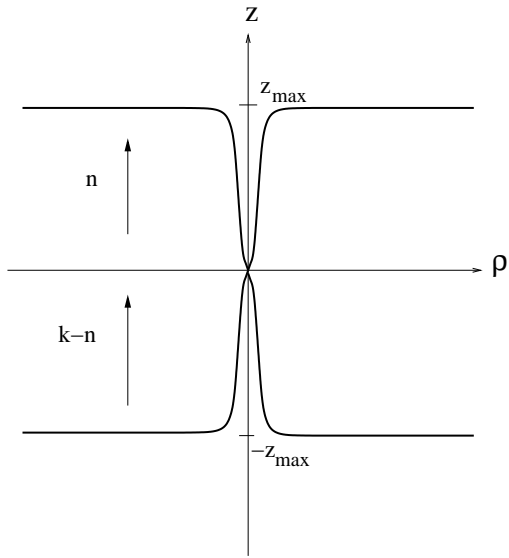}
\vspace{-0.7cm}
\end{center}
\caption{
Two D3-branes with the same shape but placed on the 
opposite side of NS5-branes along $z$ direction, which 
corresponds to pair of Cardy boundary states $\vert n>_{C}$ 
and  $ \vert k-n >_{C}$ 
}
\end{figure}
%%%%%%%%%%%%%%%%%%%%%%%%%%%%%%%%%%%%%%%%%%
\noindent 

For definiteness, we assume that D-strings extend along 
$z$ direction, that is, for the lower D3-brane, $k-n$ 
D-strings emanate from the lower flat D3-brane and terminate 
on NS5-branes, while for the upper D3-brane, $n$ D-strings 
direct away from NS5-branes and end to the upper flat D3-brane.   
When we define the D-strings emanate from NS5-brane to 
their own D3-branes, 
we see there are $n$ and $n-k$ D-strings for the upper and 
lower D3-branes\footnote{We have reversed the direction of 
the D-strings suspending between the lower flat D3-brane and 
the NS5-branes. Originally, they extend along $z$ direction and 
the number of D-strings is $k-n$, after reversion we denote the 
number of D-strings as $n-k$. } 
which is drwan in Fig. 5. 

%%%%%%%%%%%%%%%%%% Fig. 5 %%%%%%%%%%%%
\begin{figure}[htbp]
\begin{center}
\includegraphics*[scale=.40]{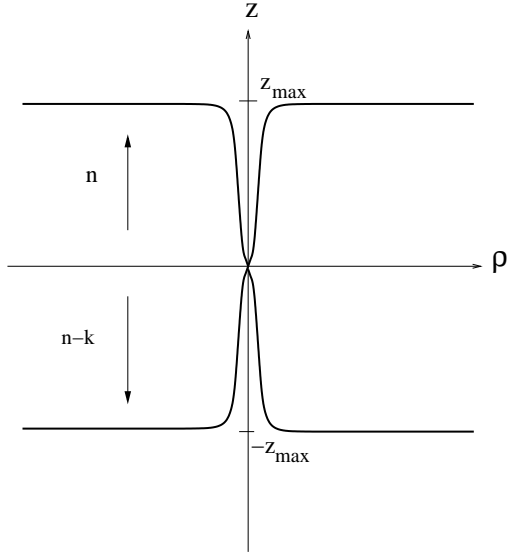}
\vspace{-0.7cm}
\end{center}
\caption{
The D-strings emanate from NS5-branes to their own 
D3-branes, and the numbers of $n$ and $n-k$ indicate the 
number of the D-strings for each type. 
}
\end{figure}
%%%%%%%%%%%%%%%%%%%%%%%%%%%%%%%%%%%%%%%%%%

As the Cardy state $\vert k-n >_{C}$ can be obtained by 
rotating $\vert n >_{C}$ by $\pi $ in the plane 
$(y^{6}, y^{7})$, similarly the lower D3-brane can be obtained 
from the upper D3-brane by the same rotation \cite{pelc}, but 
this rotating operation is topologically equivalent to moving the 
upper D3-brane across to the other side of NS5-brane along $z$ 
direction. From the correspondence between Cardy boundary state 
and the D3-brane configuration in the full asymptotically flat 
geometry of the NS5-branes, we see in Fig. 5 that  
 when the lower D3-brane passes through the $k$ coinciding 
NS5-branes, $k$ D-strings are created. In other words, the
 Hanany-Witten effect can be read off from the Cardy boundary 
states through the above correspondence, especially the number 
of the created (or annihilated) 
D-strings is completely determined by the Cardy boundary state 
$\vert n >_{C}$ and its dual $\vert k-n >_{C}$ ! 

Since the intrinsic direction of the D1-strings is defined as 
they emanate from NS5-branes to their own D3-brane, the above 
string creation  effect can be rephrased as that depending on which 
side the distant observer stays, he (she) will measure two sorts 
of the number of D-strings: either $n$ or $n-k$. 
In $SU(2)$ WZW model, the spherical D-branes can be contracted 
either to the group unit $e$ along the 3-ball $\Gamma $ or 
to the opposite pole $(-e)$ of $S^{3}_{6789}$ along the 3-ball 
$\Gamma '$ \cite{alekseev2}, \cite{figueroa}. 
Two ways of computing RR Page D0 charge of a spherical D2-brane 
result in two RR Page charge $n$ or $n-k$, and the flux is 
only determined modulo $k$ if we demand in physical process 
the RR Page D0 charge is reserved. In \cite{alekseev2}, this 
peculiar feature of the RR Page D0 charge was expected to be
related to the fact
that $H$-field belongs to a nontrivial cohomology class. 
However, in the present context, we find that two sorts of 
RR Page charge is due to two different relative position 
between the observer and the measured D3-brane. To be more 
precise, suppose that D3-brane is placed on the upper side 
of the NS5-branes, the distant observer can stay on upper side 
and the measured RR Page charge is $n$, but he (she) can 
also choose to stay on lower side and the observed RR Page D0 
charge is $n-k$, which can be interpreted as $D$-string creation 
in the full NS5-brane geometry\footnote{In D0/D8 case, the number
of the fundamental strings suspending between D0- and D8-brane is
proportional to $|z|/2z$ and dependent on how to choose
positive direction for z. Usually one defines the positive direction
for z as that it extends from D8-brane to the observer, and the
observer will measure $+1/2$ F-string if D0-brane is at the 
observer's side, but $-1/2$ F-string if at the 
opposite side \cite{shimizuzhou}.}. Actually, in the above physical 
process, RR Page D0 charge is not a conserved quantity.    
When we define RR Page D0 charge modulo $k$, we effectively 
erase D-string creation effect in 10D curved space. 
If we use brane source charge instead of Page charge, 
the Hanany-Witten effect is also removed. 

In summary, we have observed the correspondence between the 
Cardy boundary states in $SU(2)$ group manifold and the BPS 
D3-brane configuration preserving the $SO(3)_{789}$ symmetry 
in the full asymptotically flat geometry of the $k$ 
coincident NS5-branes. The dual Cardy boundary state has 
been constructed by acting the rotation operator on 
$\vert n >_{C}$, which rotates $\pi $ in $(y^{6}, y^{7})$ 
plane. Exploiting the correspondence and the dual Cardy boundary 
state, we have found that the Hanany-Witten effect
, the generalization of the string creation mechanism for the 
case of D0/D8-branes \cite{danielsson}-\cite{kitao},  
can be induced from the Cardy boundary states. By rephrasing 
 Hanany-Witten effect, we have shown that due to the 
contraction to $e$ or $(-e)$,  the two different RR Page D0 
charges for the $n$-th spherical D2-brane in $SU(2)$ WZW model 
can be interpreted as 
string creation in the background of the $k$ coinciding 
NS5-branes.

\begin{flushleft}
\bf Acknowledgements
\end{flushleft}

We would like to thank P.M. Ho and M. Kreuzer for valuable 
discussions. The work of T.K. is supported in part by Scientific Grants 
from the Ministry of Education (grant Nr., 09640353).
The work of J.-G. Z. is supported in part by the 
Austrian Research Funds FWF under grant Nr. M535-TPH.

\end{document}